\documentclass[aps,pre,twocolumn,amsmath,amssymb,showpacs]{revtex4}

\usepackage{graphicx}
\usepackage{dcolumn}
\usepackage{bm}

\begin{document}

\title{Weyl formulas for annular ray-splitting billiards}

\author{Yves D\'ecanini}
\email{decanini@univ-corse.fr}
\author{Antoine Folacci}
\email{folacci@univ-corse.fr}
\affiliation{
SPE, UMR CNRS 6134, Equipe Physique Semi-Classique (et)
de la Mati\`ere Condens\'ee \\
Universit\'e de Corse, Facult\'e des Sciences, BP 52, 20250 Corte,
France}

\date{\today}

\begin{abstract}
We consider the distribution of eigenvalues for the wave equation
in annular (electromagnetic or acoustic) ray-splitting billiards.
These systems are interesting in that the derivation of the
associated smoothed spectral counting function can be considered
as a canonical problem. This is achieved by extending a formalism
developed by Berry and Howls for ordinary (without ray-splitting)
billiards. Our results are confirmed by numerical computations and
permit us to infer a set of rules useful in order to obtain Weyl
formulas for more general ray-splitting billiards.
\end{abstract}

\pacs{05.45.Mt, 03.65.Sq, 42.25.-p, 43.20.+g}

\maketitle

\section{Introduction}

Since the pioneering work of H. Weyl \cite{Weyl1911} dealing with
the distribution of eigenvalues for the wave equation in a cavity
with a perfectly reflecting boundary, considerable effort has been
devoted to the construction of the smooth part of spectral
counting functions (Weyl formulas) in various fields of physics
and mathematics. For an historical point of view on this problem,
we refer to the seminal paper of Kac \cite{Kac} and to the
monograph of Batles and Hilf \cite{BaltesHilf}. For its importance
in physics and for various applications, we refer to the
monographs of Batles and Hilf \cite{BaltesHilf} and of Brack and
Bhaduri \cite{BrackBhaduri} as well as to references therein.

In general, the determination of the smooth part of a spectral
counting function is a complicated task and only the leading order
terms can be obtained. By contrast, for two- and three-dimensional
billiards with Dirichlet, Neumann or Robin boundary conditions,
this problem can be considered as definitely solved (see, e.g.,
Refs.~\onlinecite{BalianBloch1, BalianBloch2, SW71, BerryHowls,
Sieber95} or the monographs cited above). For ray-splitting
billiards introduced in the context of acoustic and quantum chaos
by Couchman \textit{et al} \cite{Couchman92} and extensively
studied these last years (see, e.g., Refs.~\onlinecite{Prange96,
Blumeletal1, Blumeletal2, KohlerBlumel1, KohlerBlumel2,
KohlerBlumel, Dabaghian2001, Savytskyy2001, Blumel2002}), it is
rather natural to think that the same outcome could be obtained.
Recently, some progress has been made in that direction
\cite{Prange96, Blumeletal2, KohlerBlumel1, KohlerBlumel} but it
seems we are very far from a general theory. With this aim in
view, it is interesting to solve canonical problems, i.e., to
consider simple examples of ray-splitting billiard problems for
which it is possible to perform exactly the calculations
\cite{Dabaghian2001, Blumel2002} or to carry on them as far as
possible \cite{KohlerBlumel}. The results then obtained are useful
in order to infer Weyl formulas for more general ray-splitting
billiards.

With this in mind, we are concerned, in this paper, with the
distribution of eigenvalues for the scalar wave equation in a
two-dimensional dielectric annular billiard. This billiard
consists of an outer circle with radius $R$ and an inner circle
with radius $r$. The index of refraction between the two circles
(region I) is fixed at 1 while the inner disk (region II) is
characterized by the index of refraction $N$. At the interface
between the two regions, we shall assume that the scalar field
$\Phi$ solution of the wave equation and its normal derivative
$\partial \Phi /
\partial n$ satisfy the boundary conditions
\begin{equation}
\Phi^{\mathrm I} = \Phi^{\mathrm {II}} \quad \mathrm{and} \quad
\frac{\partial \Phi^{\mathrm I}}{\partial n} =\alpha
\frac{\partial \Phi^{\mathrm {II}}}{\partial n}.
\end{equation}
The cases $\alpha = 1$ and $\alpha = 1/N^2$ respectively
correspond to the TM and TE polarizations in electromagnetism
\cite{Jones86}. On the outer circle, we shall assume that $\Phi$
vanishes (Dirichlet boundary condition). Such a condition is more
artificial than physical. It can be partially realized for the TM
polarization if the billiard is embedded in a perfect conductor.
We assume it in order to simplify our calculations. Mutatis
mutandis, we can also consider the case of a two-dimensional
acoustic annular billiard. In that case, region I (resp. region
II) is occupied by a perfect fluid with density $\rho^{\mathrm
{I}}$ (resp. $\rho^{\mathrm {II}}$) while $\alpha= \rho^{\mathrm
{I}} / \rho^{\mathrm {II}}$ \cite{Jones86} and can take any
positive value. Finally, in order to be able to analytically
perform the calculations, we shall assume that the two circles are
concentric, but the final results given by
Eqs.~\ref{NbarTM}-\ref{NbarGen} and \ref{NbarCORR} are equally
valid for non-concentric circles.

For such billiards, eigenvalues cannot be analytically obtained.
They satisfy a transcendental equation involving Bessel functions
that can be solved only numerically. In spite of this, we are able
to analytically derive the associated Weyl formulas from the
corresponding Green functions. This is done by using an approach
developed by Berry and Howls \cite{BerryHowls} and which
generalizes a previous work by Stewartson and Waechter
\cite{SW71}. This approach has been considered by these authors
for the circular billiard with the Dirichlet condition on its
boundary. We extend it rather naturally to the more complicated
case of annular ray-splitting billiards. We are then confronted by
some tedious algebraic calculations which, fortunately, can be
performed with the help of {\it Mathematica} \cite{Wolfram1996}.

Our paper is organized as follows. In Section 2, we introduce our
notations and we construct the Green function for the annular
ray-splitting billiard as well as the associated regularized
resolvent. In Section 3, by extending the Berry-Howls approach, we
obtain a set of Weyl formulas corresponding to various values of
the parameters $\alpha$ and $N$. In Section 4, we briefly consider
the same problem for the desymmetrized versions of the annular
ray-splitting billiard. In Section 5, we numerically check the
previous results and, in Section 6, we conclude our paper by
inferring a set of rules useful for constructing Weyl formulas for
more general ray-splitting billiards (see
Eqs.~\ref{GenRulesWF1}-\ref{GenRulesWF8}).

\section{Green function for the annular ray-splitting billiard
and regularized resolvent}

From now on, we shall use the polar coordinate system
$(\rho,\theta)$ with its origin $O$ at the common center of the
two circles which define the annular ray-splitting billiard. The
eigenvalues $k_i$ for the wave equation in this billiard as well
as the associated eigenfunctions $\Phi_i$ are determined by
solving the following problem:
\begin{itemize}
  \item (i) The $k_i$ and $\Phi_i$ satisfy the Helmholtz equation
\begin{equation}\label{EVP}
{\hat{H}}_{\bf x} \, \Phi_i({\bf x})=k_i^2 \, \Phi_i({\bf x})
\quad {\bf x}(\rho,\theta)
\end{equation}
where
\begin{equation}\label{Hdef}
{\hat{H}}_{\bf x}=\left\{ \begin{array}{cl} -\Delta_{\bf x} & \mathrm{for} \ r<\rho < R, \\
 -(1/N^{2})\Delta_{\bf x} & \mathrm{for} \ 0<\rho < r,
\end{array}
\right.
\end{equation}
with the Laplacian $\Delta_{\bf x}$ given, in the polar coordinate
system, by
\begin{equation}\label{lapl}
\Delta_{\bf x} = \frac{\partial ^2}{\partial
\rho^2}+\frac{1}{\rho}\frac{\partial }{\partial
\rho}+\frac{1}{\rho^2} \frac{\partial ^2}{\partial \theta^2}.
\end{equation}
  \item (ii) The $\Phi_i$ satisfy the boundary conditions
\begin{subequations}\label{BCPhi}
\begin{eqnarray}
&&\Phi_i^{\mathrm I}(\rho =R,\theta)=0 , \\
&&\Phi_i^{\mathrm I}(\rho =r,\theta)=\Phi_i^{\mathrm {II}}(\rho
=r,\theta) , \\
&&\frac{\partial \Phi_i^{\mathrm I}}{\partial \rho}(\rho
=r,\theta)=\alpha \, \frac{\partial \Phi_i^{\mathrm
{II}}}{\partial \rho}(\rho =r,\theta) ,
\end{eqnarray}
\end{subequations}
for $0\leq\theta<2\pi$.
\end{itemize}
Because a solution of (\ref{EVP}) is expressible in terms of
Bessel functions \cite{AS65}, it is easy to prove from
Eq.~\ref{BCPhi} that the $k_i$ are the values of $k$ which solve
\begin{equation} \label{detEV}
\left|
 \begin{array}{ccccc} J_m(kR) & J_m(kr) & J'_m(kr) \\
  H^{(1)}_m(kR)& H^{(1)}_m(kr) & H'^{(1)}_m(kr)   \\
  0 & J_m(Nkr) & N\alpha J'_m(Nkr)  \\
  \end{array}
 \right| =0
\end{equation}
for $m \in {\bf Z}$. For a given $m$, they can be indexed by the
integer $n$ with $n = 1, 2, 3, \dots $ and the corresponding
eigenfunctions are given by
\begin{eqnarray} \label{PhiNM}
& & \Phi_{m,n}(\rho,\theta)=A_{m,n} \biggl( J_m( k_{m,n}\rho )  \biggr. \nonumber \\
& & \qquad \qquad \quad \biggl.
-\frac{J_m(k_{m,n}R)}{H^{(1)}(k_{m,n}R)} H^{(1)}(k_{m,n}\rho)
\biggr) e^{im\theta}
\end{eqnarray}
where the $A_{m,n}$ are normalization constants. It should be
noted that the eigenvalues corresponding to $m\not=0$ are
twofold-degenerated because of the relation $k_{m,n} = k_{-m,n}$
which follows from the invariance of Eq.~\ref{detEV} under the
change $m \to -m$.

The determination of the eigenvalues $k_i$ permits us to construct
the spectral counting function associated with the annular
ray-splitting billiard. It is given by
\begin{equation}\label{FC}
{\mathcal N}(k)=\sum_i \Theta(k-k_i)=\sum_{m=-\infty}^{+\infty }
\sum_{n=1}^{+\infty }\Theta(k-k_{m,n})
\end{equation}
where $\Theta $ denotes the Heaviside function.

Eq.~\ref{detEV} can be solved only numerically. As a consequence,
the smooth part of the spectral counting function ${\mathcal
N}(k)$ cannot be obtained directly from (\ref{FC}). In order to
accomplish this, it is more convenient to generalize the
Berry-Howls approach \cite{BerryHowls} (see also Stewartson and
Waechter \cite{SW71}) by introducing the regularized resolvent
\begin{eqnarray}\label{gs}
g(s)=\int_{0}^{r}\! \int_{0}^{2\pi}\left[G^{\mathrm {II}}({\bf
x},{\bf x},s)-G_0^{\mathrm {II}}({\bf x},{\bf
x},s)\right]\,\rho\,d\rho\,d\theta \,
+\nonumber\\
\, \int_{r}^{R}\! \int_{0}^{2\pi}\left[G^{\mathrm I}({\bf x},{\bf
x},s)-G_0^{\mathrm I}({\bf x},{\bf
x},s)\right]\,\rho\,d\rho\,d\theta\quad
\end{eqnarray}
where $G_{0}$ is the ``free-space Green function" given by
\begin{subequations}
\begin{eqnarray}\label{FG}
&&G_0^{\mathrm I}({\bf x},{\bf
x'},s)=\frac{1}{2\pi}\,K_0(s\left|{\bf
x}-{\bf x'}\right|)\nonumber\\
&&\quad
=\frac{1}{2\pi}\sum_{m=-\infty}^{+\infty}I_m(s\rho_<)K_m(s\rho_>)
e^{im(\theta-\theta')}, \\
&&G_0^{\mathrm {II}}({\bf x},{\bf
x'},s)=\frac{N^2}{2\pi}\,K_0(Ns\left|{\bf
x}-{\bf x'}\right|)\nonumber\\
&&\quad
=\frac{N^2}{2\pi}\sum_{m=-\infty}^{+\infty}I_m(Ns\rho_<)K_m(Ns\rho_>)
e^{im(\theta-\theta')}, \nonumber\\
\end{eqnarray}
\end{subequations}
while $G$ is the annular ray-splitting billiard Green function
solution of \begin{equation}\label{Gdef} ({\hat{H}}_{\bf
x}+s^2)\,G({\bf x},{\bf x'},s)=\delta ({\bf x}-{\bf x'})
\end{equation}
and subject to the boundary conditions
\begin{subequations}\label{BCG}
\begin{eqnarray}
&&G^{\mathrm I}({\bf x},{\bf x'},s)=0 \nonumber\\
&&\qquad \mathrm{for}\quad \rho~\mathrm{or}~\rho'=R \quad
\mathrm{and}\quad 0\leq\theta,\,\theta'<2\pi, \\
&&G^{\mathrm I}({\bf x},{\bf x'},s)=G^{\mathrm {II}}({\bf x},{\bf x'},s)\nonumber\\
&&\qquad \mathrm{for}\quad \rho~\mathrm{or}~\rho'=r \quad
\mathrm{and}\quad 0\leq\theta,\,\theta'<2\pi,\\
&&\frac{\partial G^{\mathrm I}}{\partial \rho}({\bf x},{\bf
x'},s)=\alpha\,\frac{\partial G^{\mathrm {II}}}{\partial
\rho}({\bf
x},{\bf x'},s)\nonumber\\
&&\qquad \mathrm{for}\quad \rho~\mathrm{or}~\rho'=r \quad
\mathrm{and}\quad 0\leq\theta,\,\theta'<2\pi.
\end{eqnarray}
\end{subequations}
Here, it should be noted that in order to construct $g(s)$, we
need the Green function $G({\bf x},{\bf x'},s)$ only for ${\bf x}$
and ${\bf x'}$ lying in the same region of the billiard.

When $|s|$ is large, $g(s)$ has the asymptotic expansion (Weyl
series)
\begin{equation}\label{gsAS}
g(s)=\sum_{p=1}^{+\infty }\,\frac{c_p}{s^p}
\end{equation}
and from the $c_p$-coefficients we can obtain the large-$k$
asymptotic behavior for the spectral counting function in the form
\begin{equation}
\mathcal{N}(k)=\frac{\mathcal{A}}{4\pi}k^2+\frac{2c_1}{\pi}\,k+c_2-\frac{k}{\pi}
\sum_{n=1}^{+\infty}\frac{(-1)^n}{(n-\frac{1}{2})\,k^{2n}}\,c_{2n+1}.
\end{equation}
Our theory does not provide the expression for the surface term
$\mathcal{A}$. We shall assume that $\mathcal{A}$ is the billiard
total area weighted by the refraction index and given by
\begin{equation}\label{Aire}
\mathcal{A}=\pi(R^2-r^2)+N^2\pi r^2.
\end{equation}

In order to obtain the $c_p$-coefficients, we must first solve the
problem defined by Eqs.~\ref{Gdef} and \ref{BCG} and then to
perform the integrations in Eq.~\ref{gs}. The solution of
(\ref{Gdef}) and (\ref{BCG}) can be constructed in terms of the
modified Bessel functions \cite{AS65} and is given by
\begin{widetext}
\begin{subequations}
\begin{eqnarray}\label{GExp}
&&G^{\mathrm I}({\bf x},{\bf
x'},s)=\frac{1}{2\pi}\sum_{m=-\infty}^{+\infty} \frac{I_m(sR)
D_m^{(1)}(sr,N)}{I_m(sR) D_m(sr,N)-K_m(sR) D_m^{(1)}(sr,N)}
\nonumber  \\
&& \qquad \qquad \qquad \qquad \qquad \qquad
\left[K_m(s\rho_>)-\frac{K_m(sR)}{I_m(sR)}\,I_m(s\rho_>)\right]
\left[\frac{D_m(sr,N)}{D_m^{(1)}(sr,N)}\,I_m(s\rho_<)-K_m(s\rho_<)\right]
\,e^{im(\theta-\theta')}\\
&&G^{\mathrm {II}}({\bf x},{\bf
x'},s)=\frac{N^2}{2\pi}\sum_{m=-\infty}^{+\infty} I_m(Ns\rho_<)
\Bigg[K_m(Ns\rho_>) -\frac{I_m(sR) E_m(sr,N)-K_m(sR)
E_m^{(1)}(sr,N)}{I_m(sR) D_m(sr,N) -K_m(sR) D_m^{(1)}(sr,N)}
I_m(Ns\rho_>)\Bigg] e^{im(\theta-\theta')} \nonumber\\
\end{eqnarray}
\end{subequations}
with $\rho_<=\mathrm{inf}(\rho,\rho'),
\rho_>=\mathrm{sup}(\rho,\rho')$ and
\begin{subequations}
\begin{eqnarray} \label{DmEm}
&&D_m(sr,N)=K'_m(sr)\,I_m(Nsr)-N\alpha\, I'_m(Nsr)\,K_m(sr),\\
&&D_m^{(1)}(sr,N)=I'_m(sr)\,I_m(Nsr)-N\alpha\,I'_m(Nsr)\,I_m(sr),\\
&&E_m(sr,N)=K'_m(sr)\,K_m(Nsr)-N\alpha\,K'_m(Nsr)\,K_m(sr), \\
&&E_m^{(1)}(sr,N)=I'_m(sr)\,K_m(Nsr)-N\alpha\,K'_m(Nsr)\,I_m(sr).
\end{eqnarray}
\end{subequations}
This provides the expression for $g(s)$:
\begin{equation}\label{gsnew}
g(s)=-\frac{1}{2}\sum_{m=-\infty}^{+\infty}f_m(s)
\end{equation}
where
\begin{equation}\label{fms}
f_m(s)=f_m^{(1)}(s)+f_m^{(2)}(s)+f_m^{(3)}(s)
\end{equation}
with
\begin{subequations}\label{fms123}
\begin{eqnarray}
&&f_m^{(1)}(s)=R^2\left[\left(1+\frac{m^2}{s^2R^2}\right)\,I_m(sR)\,K_m(sR)
-I'_m(sR)\,K'_m(sR)
\right.
\nonumber \\
&& \qquad \qquad -  \left. \frac{1}{sR}\
\frac{I'_m(sR)\,D_m(sr,N)-K'_m(sR)\,D_m^{(1)}(sr,N)}{I_m(sR)\,D_m(sr,N)-K_m(sR)
\,D_m^{(1)}(sr,N)}\right],
\\
&&f_m^{(2)}(s)=-r^2\left[\left(1+\frac{m^2}{s^2r^2}\right)\,I_m(sr)\,K_m(sr)
-I'_m(sr)
\,K'_m(sr) \right.\nonumber \\
&& \qquad \qquad -\frac{N\alpha}{sr}\
\frac{I_m(sR)\,K'_m(sr)-K_m(sR)\,I'_m(sr)}{I_m(sR)
\,D_m(sr,N)-K_m(sR)\,D_m^{(1)}(sr,N)}
\,I'_m(Nsr)\nonumber\\
&& \qquad \qquad  \left.
+\left(1+\frac{m^2}{s^2r^2}\right)\,\frac{1}{sr}\
\frac{I_m(sR)\,K_m(sr)-K_m(sR)\,I_m(sr)}{
I_m(sR)\,D_m(sr,N)-K_m(sR)\,D_m^{(1)}(sr,N)} \,I_m(Nsr)\right],
\end{eqnarray}
\begin{eqnarray}
&&f_m^{(3)}(s)=N^2r^2\left[\left(1+\frac{m^2}{N^2s^2r^2}\right)\,I_m(Nsr)\,K_m(Nsr)
-I'_m(Nsr)\,K'_m(Nsr)\right.\nonumber \\
&& \qquad \qquad -\frac{1}{Nsr}\
\frac{I_m(sR)\,K'_m(sr)-K_m(sR)\,I'_m(sr)}{I_m(sR)\,D_m(sr,N)-
K_m(sR)\,D_m^{(1)}(sr,N)}
\,I'_m(Nsr) \nonumber\\
&& \qquad \qquad \left. +\left(1+\frac{m^2}{N^2s^2r^2}\right)\,
\frac{\alpha}{sr}\ \frac{I_m(sR)\,K_m(sr)-K_m(sR)\,I_m(sr)}{
I_m(sR)\,D_m(sr,N)-K_m(sR)\,D_m^{(1)}(sr,N)} \,I_m(Nsr)\right].
\end{eqnarray}
\end{subequations}
\end{widetext}
By using the Poisson summation formula as well as the relation
$f_{-m}(s)=f_m(s)$, we can write
\begin{equation}\label{Gpoi}
g(s)=-\sum_{\mu=-\infty}^{+\infty} \int_{0}^{+\infty}f_m(s)
e^{i2\pi\mu m} \ dm.
\end{equation}
It should be noted that (\ref{Gpoi}) with $f_m(s)$ given by
(\ref{fms}) and (\ref{fms123}) provides an exact expression for
$g(s)$.

\section{From the regularized resolvent to the smoothed spectral
counting function}

The large-$|s|$ asymptotic behavior (\ref{gsAS}) of $g(s)$ can now
be found from (\ref{Gpoi}) by replacing in Eqs.~\ref{fms} and
\ref{fms123} the modified Bessel functions $I_m, I'_m, K_m$ and
$K'_m$ by their uniform asymptotic expansions (see Eqs.~9.7.8 -
9.7.10 of Ref.~\onlinecite{AS65}) given by
\begin{subequations}\label{UAsymp}
\begin{eqnarray}
&&I_m(z)=\frac{1}{\sqrt{2\pi}} \frac{1}{(m^2+z^2)^{\frac{1}{4}}}\
\mathrm{exp}(F_m(z)/2)\nonumber\\
&&\qquad \qquad \qquad \qquad \times
\sum_{p=0}^{+\infty}\frac{u_p\left[t_m(z)\right]}{m^p},\\
&&I'_m(z)=\frac{(m^2+z^2)^{\frac{1}{4}}}{z\sqrt{2\pi}}\
\mathrm{exp}(F_m(z)/2)\nonumber\\
&&\qquad \qquad \qquad \qquad \times
\sum_{p=0}^{+\infty}\frac{v_p\left[t_m(z)\right]}{m^p},\\
&&K_m(z)=\sqrt{\frac{\pi}{2}} \frac{1}{(m^2+z^2)^{\frac{1}{4}}}\
\mathrm{exp}(-F_m(z)/2)\nonumber\\
&&\qquad \qquad \qquad \qquad \times
\sum_{p=0}^{+\infty}\frac{(-1)^pu_p\left[t_m(z)\right]}{m^p},\\
&&K'_m(z)=-\sqrt{\frac{\pi}{2}} \frac{(m^2+z^2)^{\frac{1}{4}}}{z}\
\mathrm{exp}(-F_m(z)/2)\nonumber\\
&&\qquad \qquad \qquad \qquad \times
\sum_{p=0}^{+\infty}\frac{(-1)^pv_p\left[t_m(z)\right]}{m^p}.
\end{eqnarray}
\end{subequations}
Here
\begin{subequations}
\begin{eqnarray}
&&F_m(z)=2(m^2+z^2)^{\frac{1}{2}}+2m\
\log\left[\frac{z}{m+(m^2+z^2)^{\frac{1}{2}}}\right],\nonumber\\
\\
&&t_m(z)=\frac{m}{(m^2+z^2)^{\frac{1}{2}}},
\end{eqnarray}
\end{subequations}
and $u_p$ and $v_p$ are polynomials given in chapter 9 of
Ref.~\onlinecite{AS65} (Eqs.~9.3.10 and 9.3.14). It should be
noted that, as for the circle billiard \cite{SW71, BerryHowls},
the Weyl coefficients $c_p$ and therefore the smoothed spectral
counting function come directly from the $\mu=0$-term in
(\ref{Gpoi}). As noted by Berry and Howls \cite{BerryHowls} (see
also \cite{HowlsTrasler1, HowlsTrasler2}), the other terms are
associated with the fluctuating part of the spectral counting
function which could be obtained by carefully taking into account
Stokes phenomenon for the asymptotic expansions (\ref{UAsymp}) in
the context of hyperasymptotics \cite{hyperasymp1, hyperasymp2}.

The $\mu=0$-term in (\ref{Gpoi}) now reduces to
\begin{equation}\label{gbar1}
\bar{g}(s)=-\int_{0}^{+\infty}\bar{f}_m(s) \ dm.
\end{equation}
where
\begin{equation}\label{gbar2}
\bar{f}_m(s)=\bar{f}_m^{(1)}(s)+\bar{f}_m^{(2)}(s)+\bar{f}_m^{(3)}(s)
\end{equation}
with
\begin{subequations}\label{fbar}
\begin{eqnarray}
&&\bar{f}_m^{(1)}(s)=-\frac{\sqrt{m^2+s^2R^2}}{s^2}
\sum_{p=1}^{+\infty}
\frac{A_p^{(1)}\left[t_m(sR)\right]}{m^p},\\
&&\bar{f}_m^{(2)}(s)=+\frac{\sqrt{m^2+s^2r^2}}{s^2}
\sum_{p=1}^{+\infty}
\frac{A_p^{(2)}\left[t_m(sr),t_m(Nsr)\right]}{m^p}, \nonumber\\
\\
&&\bar{f}_m^{(3)}(s)=-\frac{\sqrt{m^2+N^2s^2r^2}}{s^2}
\sum_{p=1}^{+\infty}
\frac{A_p^{(3)}\left[t_m(sr),t_m(Nsr)\right]}{m^p}.\nonumber\\
\end{eqnarray}
\end{subequations}
Here the functions $A_p^{(1)}$, $A_p^{(2)}$ and $A_p^{(3)}$ can be
expressed in terms of the polynomials $u_p$ and $v_p$ and
therefore can be explicitly obtained (see, below, Eq.~\ref{AAA}).
Then, by using Eqs.~\ref{gbar1}-\ref{fbar}, we find the general
expression
\begin{widetext}
\begin{eqnarray}
c_p&=&\frac{1}{R\,^{p-2}}\int_{0}^{+\infty}\frac{\sqrt{1+x^2}}{x\,^p}\,
A_p^{(1)}\left[x/\sqrt{1+x^2}\right]\,dx \nonumber\\
& & \qquad \qquad -\frac{1}{r\,^{p-2}}\,
\left[\int_{0}^{+\infty}\frac{\sqrt{1+x^2}}{x\,^p}\,A_p^{(2)}
\left[x/\sqrt{1+x^2},x/\sqrt{N^2+x^2}\right]\,dx  \right.  \nonumber\\
& & \qquad \qquad  \qquad \qquad
\left.-\int_{0}^{+\infty}\frac{\sqrt{N^2+x^2}}{x\,^p}\,
A_p^{(3)}\left[x/\sqrt{1+x^2},x/\sqrt{N^2+x^2}\right]\,dx \right].
\end{eqnarray}

In order to provide the terms in $k$ and $k^0$ in the expression
of the smoothed spectral counting function, we need the
coefficients $c_1$ and $c_2$. They can be obtained, by performing
the integrations in the previous equation, from the functions
$A_1^{(1)}$, $A_2^{(1)}$, $A_1^{(2)}$, $A_2^{(2)}$, $A_1^{(3)}$
and $ A_2^{(3)}$ which are explicitly given by
\begin{subequations}\label{AAA}
\begin{eqnarray}
&&A_1^{(1)}\left[x/\sqrt{1+x^2}\right]=-\frac{x}{2\,(1+x^2)^{3/2}}, \\
&&A_2^{(1)}\left[x/\sqrt{1+x^2}\right]=\frac{x^4}{2\,(1+x^2)^3},  \\
&&A_1^{(2)}\left[x/\sqrt{1+x^2},x/\sqrt{N^2+x^2}\right]=
\frac{x}{2\,(1+x^2)^{3/2}}-\frac{x}{(1+x^2)\,
\left[\sqrt{1+x^2}+\alpha\,\sqrt{N^2+x^2}\right]},   \\
&&A_2^{(2)}\left[x/\sqrt{1+x^2},x/\sqrt{N^2+x^2}\right]=
\frac{\left[(1-\alpha)N^2+(1-\alpha N^2)^2x^2+
(\alpha-1)(2\alpha+1)N^2x^4 + (\alpha^2-1)x^6 \right]x^2}
{2(N^2+x^2)(1+x^2)^3\left[\sqrt{1+x^2}+\alpha\,
\sqrt{N^2+x^2}\right]^2},\\
&&A_1^{(3)}\left[x/\sqrt{1+x^2},x/\sqrt{N^2+x^2}\right]=
-\frac{N^2 \, x}{2\,(N^2+x^2)^{3/2}}+
\frac{\alpha\,N^2\,x}{(N^2+x^2)\,\left[\sqrt{1+x^2}+\alpha\,
\sqrt{N^2+x^2}\right]},  \\
&&A_2^{(3)}\left[x/\sqrt{1+x^2},x/\sqrt{N^2+x^2}\right]=
\frac{\left[\alpha (\alpha-1)N^6 + (1-\alpha N^2)^2 N^2 x^2
+(1-\alpha)(\alpha+2)N^2 x^4  +(1-\alpha^2) N^2 x^6\right]x^2}
{2(1+x^2)(N^2+x^2)^3
\left[\sqrt{1+x^2}+\alpha\,\sqrt{N^2+x^2}\right]^2}.  \nonumber\\
\end{eqnarray}
\end{subequations}
For the TM polarization ($\alpha=1$), we then find
\begin{subequations}\label{NbarTM}
\begin{eqnarray}
&&\bar {\mathcal{N}}(k) =
\frac{k^2(R^2-r^2)}{4}+\frac{N^2k^2r^2}{4}
-\frac{kR}{2}+\left[\frac{2}{\pi}\,E(1-N^2)-\frac{N}{2}-\frac{1}{2}\right]kr+
\frac{1}{6}+
\cdots~\mathrm{if}~N<1,     \\
&&\bar {\mathcal{N}}(k) =
\frac{k^2(R^2-r^2)}{4}+\frac{N^2k^2r^2}{4}-\frac{kR}{2}+
\left[\frac{2\,N}{\pi}\,E \left(\frac{N^2-1}{N^2} \right)
-\frac{N}{2}-\frac{1}{2}\right]kr+ \frac{1}{6} +
\cdots~\mathrm{if}~N>1.
\end{eqnarray}
\end{subequations}
For the TE polarization ($\alpha=1/N^2$), we then obtain
\begin{subequations}\label{NbarTE}
\begin{eqnarray}
&& \bar {\mathcal{N}}(k) =
\frac{k^2(R^2-r^2)}{4}+\frac{N^2k^2r^2}{4}-\frac{kR}{2} \nonumber
\\
&& \quad \qquad +
\left[\frac{2}{\pi}\,\frac{N^2}{1+N^2}\,K(1-N^2)+
\frac{2}{\pi}\,\frac{N^4}{1+N^2}\,\Pi
\left(1-N^4,\frac{\pi}{2},1-N^2 \right)
-\frac{N}{2}-\frac{1}{2}\right]kr+\frac{1}{6} +
\cdots~\mathrm{if}~N<1, \\
&& \bar {\mathcal{N}}(k) =
\frac{k^2(R^2-r^2)}{4}+\frac{N^2k^2r^2}{4}-\frac{kR}{2} \nonumber
\\
&&  \quad \qquad + \left[\frac{2}{\pi}\,\frac{N}{1+N^2}\,K
\left(\frac{N^2-1}{N^2} \right)+
\frac{2}{\pi}\,\frac{1}{N(1+N^2)}\,\Pi
\left(\frac{N^4-1}{N^4},\frac{\pi}{2},\frac{N^2-1}{N^2} \right)
-\frac{N}{2}-\frac{1}{2}\right]kr+\frac{1}{6} + +
\cdots~\mathrm{if}~N>1.  \nonumber \\
&&
\end{eqnarray}
\end{subequations}
Finally, in the general case ($\alpha\not=1$), we obtain
\begin{subequations}\label{NbarGen}
\begin{eqnarray}
&& \bar {\mathcal{N}}(k) =
\frac{k^2(R^2-r^2)}{4}+\frac{N^2k^2r^2}{4}-\frac{kR}{2} \nonumber \\
&&\quad \qquad
+\left[\frac{2}{\pi}\,\frac{\alpha(1-N^2)}{\alpha^2-1}\,K(1-N^2)+
\frac{2}{\pi}\,\frac{\alpha^2N^2-1}{\alpha(\alpha^2-1)}\, \Pi
\left(\frac{\alpha^2-1}{\alpha^2},\frac{\pi}{2},1-N^2 \right)
-\frac{N}{2} -\frac{1}{2}\right]kr+\frac{1}{6} +
\cdots~\mathrm{if}~N<1, \nonumber \\
&&    \\
&& \bar {\mathcal{N}}(k) =
\frac{k^2(R^2-r^2)}{4}+\frac{N^2k^2r^2}{4}-\frac{kR}{2} \nonumber\\
&&\quad \qquad
+\left[\frac{2}{\pi}\,\frac{\alpha(1-N^2)}{N(\alpha^2-1)}\,
K\left(\frac{N^2-1}{N^2}\right)+
\frac{2}{\pi}\,\frac{\alpha(\alpha^2N^2-1)}{N(\alpha^2-1)}\,
\Pi\left(1-\alpha^2,\frac{\pi}{2},\frac{N^2-1}{N^2}\right)
-\frac{N}{2}-\frac{1}{2}\right]kr+\frac{1}{6}
+ \cdots~\mathrm{if}~N>1. \nonumber \\
\end{eqnarray}
\end{subequations}
\end{widetext}
In order to perform the integrations leading to
Eqs.~\ref{NbarTM}-\ref{NbarGen}, it has been necessary to separate
the particular case $\alpha=1$ (TM polarization) from the general
one $\alpha \not= 1$. It should be noted that the case
$\alpha=1/N^2$ (TE polarization) is included in the general case
$\alpha \not= 1$ and can be recovered from Eq.~\ref{NbarGen}.
Moreover, it is important to keep in mind that, as far as the
elliptic integrals $E$, $K$ and $\Pi$ are concerned, we adhere
with the definitions and conventions of Ref.~\onlinecite{AS65}
which are in agreement with those of {\it Mathematica}
\cite{Wolfram1996} but differ to those of Ref.~\onlinecite{GR5ed}.

The various terms appearing in Eqs.~\ref{NbarTM}-\ref{NbarGen}
have their usual physical interpretations: The first and the
second terms (in $k^2$) yield the area contributions, the third
and the fourth ones (in $k$) yield the perimeter contributions
while the fifth term (in $k^0$) yields the curvature
contributions. In particular, it should be noted that the fourth
term, in all these equations, provides the perimeter correction
associated with the circular ray-splitting boundary at $\rho = r$
and that it is this term which contains the elliptic integrals.
Moreover, it seems to us necessary to point out that the fifth
term is associated with the curvature of the circular Dirichlet
boundary at $\rho = R$. In other words, and this is rather
surprising, the ray-splitting boundary at $\rho = r$ does not
provide any correction to the curvature contributions.

\section{The desymmetrized annular ray-splitting billiard}

In Section 2, we pointed out the twofold degeneracy of the
eigenvalues $k_{m,n}$ with $m\not=0$. It is possible to work with
the non-degenerated spectra by separating the eigenfunctions of
the annular ray-splitting billiard in two different sets: In the
first set, we consider the even eigenfunctions (even in the change
${\bf x}(\rho,\theta) \to \sigma{\bf x}(\rho,-\theta)$) given by
\begin{eqnarray} \label{PhiNMe}
& & \Phi^{(+)}_{m,n}(\rho,\theta)=A^{(+)}_{m,n} \biggl( J_m(
k_{m,n}\rho )
\biggr. \nonumber \\
& & \qquad \quad \biggl. -\frac{J_m(k_{m,n}R)}{H^{(1)}(k_{m,n}R)}
H^{(1)}(k_{m,n}\rho) \biggr) \cos (m\theta)
\end{eqnarray}
with $m\in {\bf N}$ while, in the second set, we consider the odd
eigenfunctions (odd in the change ${\bf x}(\rho,\theta) \to
\sigma{\bf x}(\rho,-\theta)$) given by
\begin{eqnarray} \label{PhiNMo}
& & \Phi^{(-)}_{m,n}(\rho,\theta)=A^{(-)}_{m,n} \biggl( J_m(
k_{m,n}\rho )  \biggr.
\nonumber \\
& &  \qquad \quad \biggl. -\frac{J_m(k_{m,n}R)}{H^{(1)}(k_{m,n}R)}
H^{(1)}(k_{m,n}\rho) \biggr) \sin (m\theta)
\end{eqnarray}
with $m\in {\bf N}^*$. Here and in the following the superscripts
$(+)$ and $(-)$ refer respectively to positive and negative
parities and the $A^{(+)}_{m,n}$ and the $A^{(-)}_{m,n}$ are
normalization constants. The spectral counting functions
${\mathcal N}^{(+)}(k)$ and ${\mathcal N}^{(-)}(k)$ associated
with these two sets are then given by
\begin{subequations}\label{FCeo}
\begin{eqnarray}
& & {\mathcal N}^{(+)}(k)=\sum_{m=0}^{+\infty }
\sum_{n=1}^{+\infty
}\Theta(k-k_{m,n}),  \\
& & {\mathcal N}^{(-)}(k)=\sum_{m=1}^{+\infty }
\sum_{n=1}^{+\infty }\Theta(k-k_{m,n}).
\end{eqnarray}
\end{subequations}

The smoothed spectral counting functions ${\mathcal {\bar
N}}^{(+)}(k)$ and ${\mathcal {\bar N}}^{(-)}(k)$ respectively
associated with ${\mathcal N}^{(+)}(k)$ and ${\mathcal
N}^{(-)}(k)$ can now be obtained by using, mutatis mutandis, the
theoretical framework developed in the two previous sections:
${\mathcal {\bar N}}^{(+)}(k)$ can be constructed from the even
part $g^{(+)}(s)$ of the regularized resolvent $g(s)$ given by
(\ref{gs}) while ${\mathcal {\bar N}}^{(-)}(k)$ can be constructed
from its odd part $g^{(-)}(s)$. The functions $g^{(+)}(s)$ and
$g^{(-)}(s)$ are given by
\begin{eqnarray} \label{gseo}
g^{(\pm)}(s)&=&\frac{1}{2}\int_{0}^{r}
\int_{0}^{2\pi}\left[G^{\mathrm {II}}({\bf x},{\bf x},s)\pm
G^{\mathrm
{II}}({\bf x},\sigma{\bf x},s) \right. \nonumber  \\
& & \qquad \qquad   \qquad- \left. G_0^{\mathrm {II}}({\bf x},{\bf
x},s)\right] \,\rho\,d\rho\,d\theta \nonumber \\
&+& \frac{1}{2}\int_{r}^{R} \int_{0}^{2\pi}\left[G^{\mathrm
I}({\bf x},{\bf x},s)
 \pm G^{\mathrm I}({\bf
x},\sigma{\bf x},s) \right. \nonumber  \\
& & \qquad \qquad \qquad  - \left. G_0^{\mathrm I}({\bf x},{\bf
x},s)\right]\,\rho\,d\rho\,d\theta
\end{eqnarray}
and they satisfy
\begin{equation} \label{gsSUMgseo}
g(s)=g^{(+)}(s)+g^{(-)}(s).
\end{equation}
By performing the integrations in (\ref{gseo}), we obtain
\begin{equation}
g^{(\pm)}(s)= \frac{1}{2}g(s) \pm g^{\mathrm{corr}}(s)
\end{equation}
with
\begin{eqnarray}
& & g^{\mathrm{corr}}(s)=-\frac{1}{4}f_0(s) \nonumber \\
& & \qquad + \frac{N^2r^2}{4}\left[I_0(Nsr)K_0(Nsr)-
I'_0(Nsr)K'_0(Nsr)\right]\nonumber\\
& & \qquad + \frac{R^2}{4}\left[I_0(sR)K_0(sR)-
I'_0(sR)K'_0(sR)\right] \nonumber\\
& & \qquad -
\frac{r^2}{4}\left[I_0(sr)K_0(sr)-I'_0(sr)K'_0(sr)\right]
\end{eqnarray}
and, by using the asymptotic expansions given by Eqs.~9.7.1 -
9.7.6 of Ref.~\onlinecite{AS65}, we can write
\begin{equation}
g^{\mathrm{corr}}(s)= \frac{(N-1)r + R}{4s} - \frac{1}{8s^2} +
\underset{{|s| \to +\infty}}{\mathcal O} \left( \frac{1}{s^3}
\right).
\end{equation}
We then immediately obtain
\begin{equation} \label{NbarCORR}
{\mathcal {\bar N}} ^{(\pm)}(k) = \frac{1}{2}  \bar {\mathcal
N}(k) \pm \frac{(N-1)r+R}{2\pi}k \mp \frac{1}{8} + \cdots
\end{equation}
with $\bar {\mathcal N}(k)$ which is given by any of
Eqs.~\ref{NbarTM}-\ref{NbarGen}, according to the physical problem
considered.

Now, we would like to provide a physical interpretation of the
results obtained above. We first note that the twofold degeneracy
of the eigenvalues $k_{m,n}$ with $m\not=0$ is directly linked to
the invariance of the annular billiard under the continuous group
$O(2)$ (i.e., under rotations about the common center of the two
circles defining the billiard) and is mathematically explained by
the following result: The functions $\exp(\pm im\theta)$, with $m
\in {\bf N}^*$ fixed, form a basis for a two-dimensional
representation of $O(2)$. In order to suppress that degeneracy, it
is necessary to break the symmetry under the continuous group
$O(2)$. This can be done by folding the annular billiard along its
diameter lying on the $Ox$ axis. On that diameter, we can assume
that the scalar field $\Phi$ satisfies either the Dirichlet or the
Neumann boundary condition. We then define two different
half-annular billiards which are both desymmetrized versions of
the annular ray-splitting billiard. By assuming that the modes
which solve the problem defined by Eqs.~\ref{EVP}-\ref{BCPhi}
satisfy also the Neumann (respectively the Dirichlet) boundary
condition on the diameter, we recover the even eigenfunctions
(\ref{PhiNMe}) (respectively the odd eigenfunctions
(\ref{PhiNMo})) as well as the associated eigenvalue spectrum.
Eq.~\ref{NbarCORR} provides the Weyl formulas corresponding to
these half-annular ray-splitting billiards. The factor $1/2$ in
front of the first term of (\ref{NbarCORR}) as well as the second
and third terms are corrections which take into account the
folding of the annular billiard and the boundary conditions on the
fold. It is interesting to note i) the perimeter contribution
given by $\pm (Nr/2\pi)k$ that corresponds to the inner
half-circle diameter which bounds the region of index $N$ by a
Neumann or a Dirichlet boundary,  ii) the term $\pm 1/8$ which
originates from the two corners at the ends of the outer
half-circle diameter and iii) the fact that the ray-splitting
corners at the ends of the inner half-circle diameter do not
provide any corrections.

\section{Numerical checks}

We have checked Eqs.~\ref{NbarTM}-\ref{NbarGen} and \ref{NbarCORR}
for various configurations corresponding to different values of
the parameters $r/R$, $N$ and $\alpha$ by considering the
oscillations around zero of the function $\Delta {\mathcal {\bar
N}}(k)= {\mathcal {N}}(k) -{\mathcal {\bar N}}(k)$ which is the
fluctuating part of the spectral counting function. All these
numerical checks confirm the Weyl formulas obtained in Sections 3
and 4.

\begin{figure}
\includegraphics[height=8.0cm,width=12.0cm]{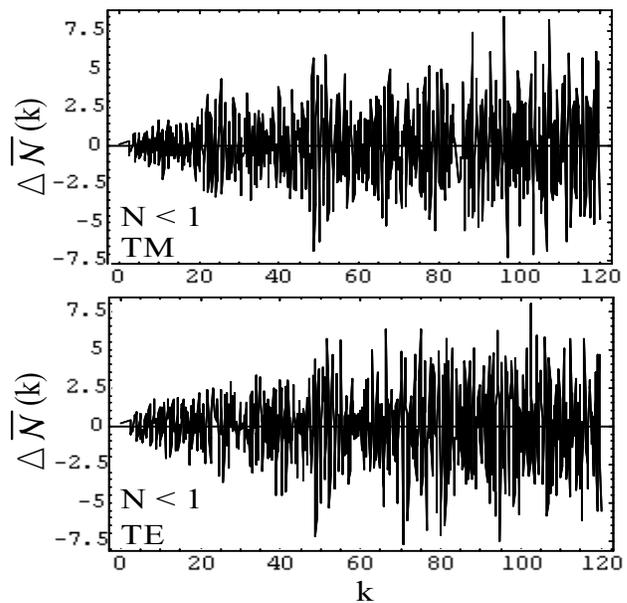}
\caption{\label{fig:Ninf1} The fluctuating part of the spectral
counting function for the TM and TE theories ($r/R=2/10$ and
$N=3/4$).}
\end{figure}

\begin{figure}
\includegraphics[height=8.0cm,width=11.0cm]{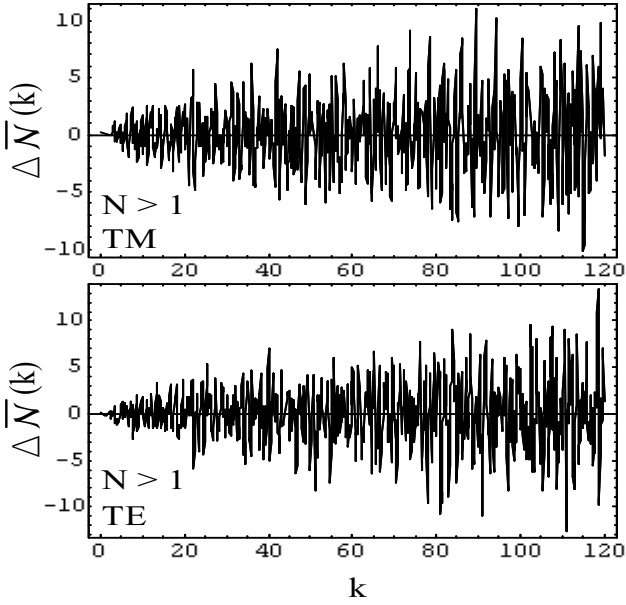}
\caption{\label{fig:Nsup1} The fluctuating part of the spectral
counting function for the TM and TE theories ($r/R=2/10$ and
$N=3$).}
\end{figure}

In Figs.~\ref{fig:Ninf1} and \ref{fig:Nsup1}, we present some
results for the TM and TE theories which have been obtained for
$r/R=2/10$ and for $N=3/4$ and $N=3$. We have computed all the
eigenvalues $k_{m,n}$ up to the frequency $k_{\mathrm{max}}=120$
by solving Eq.~\ref{detEV}. For $N=3$, the corresponding number of
eigenvalues is around $4700$ while, for $N=3/4$, it is around
3500.

\section{Conclusion and perspectives}

  a) From our previous results obtained in the particular
case of annular ray-splitting billiards, we can infer rules
permitting us to construct Weyl formulas for general ray-splitting
billiards. The associated Weyl formulas providing the smoothed
spectral counting functions are given in the usual form, i.e., by
\begin{equation} \label{GenRulesWF1}
\bar {\mathcal{N}}(k) \approx \frac{\mathcal{A}}{4\pi}k^2+
\frac{\mathcal{L}}{4\pi}k + \mathcal{C},
\end{equation}
but now, the area term $\mathcal{A}$, the perimeter term
$\mathcal{L}$ as well as the constant term $\mathcal{C}$ must take
into account the ray-splitting phenomenon:

\qquad (i) $\mathcal{A}$ is the billiard total area weighted by
the refraction index. For example, a piece of billiard of area $a$
and index of refraction $N$ provides a contribution to
$\mathcal{A}$ given by
\begin{equation}\label{GenRulesWF2}
N^2 a.
\end{equation}

\begin{widetext}
\qquad (ii) $\mathcal{L}$ is a sum of terms associated
  with the boundaries on which discontinuities in the physical
  properties occur. The contribution of a boundary of length
  $\ell$ which separates a region of index $N$ from a forbidden region
  is given by
\begin{equation}\label{GenRulesWF3}
-N\ell
\end{equation}
if we assume Dirichlet condition on that boundary and by
\begin{equation} \label{GenRulesWF4}
N\ell
\end{equation}
if we assume Neumann condition. The contribution of a boundary of
length $\ell$ which separates a region of index $N$ from a region
of index $1$ is given by
\begin{subequations} \label{GenRulesWF5}
\begin{eqnarray}
&&\left[\frac{4}{\pi} \, E(1-N^2)-N-1\right]\ell\quad\mathrm{if} \quad N<1, \\
&&\left[\frac{4N}{\pi}\,E\left(\frac{N^2-1}{N^2}\right)-N-1\right]
\ell\quad\mathrm{if} \quad N>1,
\end{eqnarray}
\end{subequations}
for the TM polarization, by
\begin{subequations}  \label{GenRulesWF6}
\begin{eqnarray}
&&\left[\frac{4}{\pi}\,\frac{N^2}{1+N^2}\,K(1-N^2)+
\frac{4}{\pi}\,\frac{N^4}{1+N^2}\,\Pi
\left( 1-N^4,\frac{\pi}{2},1-N^2 \right)
-N-1\right]\ell\quad\mathrm{if} \quad N<1,  \\
&&\left[\frac{4}{\pi}\,\frac{N}{1+N^2}\,K\left( \frac{N^2-1}{N^2}
\right)+ \frac{4}{\pi}\,\frac{1}{N(1+N^2)}\,\Pi \left
(\frac{N^4-1}{N^4},\frac{\pi}{2},\frac{N^2-1}{N^2} \right)
-N-1\right] \ell\quad\mathrm{if} \quad  N>1,
\end{eqnarray}
\end{subequations}
for the TE polarization and by
\begin{subequations}  \label{GenRulesWF7}
\begin{eqnarray}
&&\left[\frac{4}{\pi}\,\frac{\alpha(1-N^2)}{\alpha^2-1}\,K(1-N^2)+
\frac{4}{\pi}\,\frac{\alpha^2N^2-1}{\alpha(\alpha^2-1)}\,
\Pi\left(\frac{\alpha^2-1}{\alpha^2},\frac{\pi}{2},1-N^2 \right)
-N-1\right]\ell \quad\mathrm{if} \quad  N<1, \\
&&\left[\frac{4}{\pi}\,\frac{\alpha(1-N^2)}{N(\alpha^2-1)}\,K
\left(\frac{N^2-1}{N^2}\right)
+\frac{4}{\pi}\,\frac{\alpha(\alpha^2N^2-1)}{N(\alpha^2-1)}\,
\Pi\left(1-\alpha^2,\frac{\pi}{2},\frac{N^2-1}{N^2}\right)
-N-1\right]\ell\quad\mathrm{if} \quad  N>1,
\end{eqnarray}
\end{subequations}
in the general case ($\alpha\not=1$).
\end{widetext}

\qquad (iii) The constant term $\mathcal{C}$ takes into account
curvature and corner contributions. As far as the former is
concerned, it is given by
\begin{equation}  \label{GenRulesWF8}
+\frac{1}{12\pi} \int_{\Gamma} \frac{ds}{R(s)}
\end{equation}
for a boundary curve $\Gamma$ which separates a region of index
$N$ from a forbidden region, $R(s)$ denoting the local radius of
curvature along $\Gamma$. When $\Gamma$ is a ray-splitting
boundary which separates a region of index $N$ from a region of
index $1$ the associated curvature contribution vanishes. As far
as corner contributions are concerned, we simply note that
ray-splitting corners with angle $\pi/2$ provide a vanishing
contribution.

We are just beginning to check the previous formulas for the
various desymmetrized versions of ray-splitting sinai billiards.
We obtain a very good agreement between the theoretical formulas
and the numerical data. This reinforces our opinion that they are
exact. It would be very interesting to prove them rigorously but
we are unable to do so. We have also tried to link the formulas
found for the TM polarization ($\alpha=1$) to the results that
Kohler and Bl{\"u}mel \cite{KohlerBlumel} have obtained for the
scaled states of quantum ray-splitting billiards. We believe that
such a link must exist but, unfortunately, we have not established
it.

 b) In this paper, we have been exclusively concerned with the
smooth part of the spectral counting function ${\cal N}(k)$ for
annular ray-splitting billiards. It seems to us possible to treat
also the construction of the oscillating part of ${\cal N}(k)$ as
a canonical problem.  By carefully taking into account Stokes
phenomenon in the context of hyperasymptotics \cite{hyperasymp1,
hyperasymp2}, it might be possible to extract from $g(s)$ all the
periodic orbit contributions, even though the algebraic
calculations involved are certainly enormous.

 c) Finally, it would be very interesting to extend our calculations
to the three-dimensional case, having in mind applications to the
domain of quantum optics and more particularly to cavity quantum
electrodynamics. Indeed, as it is well-known, the optical
properties (spontaneous emission, stimulated emission, ... ) of
atoms and molecules embedded in a cavity strongly depend on the
density of states of the electromagnetic field. Because of that,
Weyl formulas for cavities containing dielectric structures would
be certainly welcome.

\begin{acknowledgments}
We would like to thank Bruce Jensen for discussions concerning the
article of Kac \cite{Kac} ten years ago as well as for more recent
comments on the present work.
\end{acknowledgments}

%\newpage
\bibliography{BillAnC}%

\begin{thebibliography}{27}
\expandafter\ifx\csname natexlab\endcsname\relax\def\natexlab#1{#1}\fi
\expandafter\ifx\csname bibnamefont\endcsname\relax
  \def\bibnamefont#1{#1}\fi
\expandafter\ifx\csname bibfnamefont\endcsname\relax
  \def\bibfnamefont#1{#1}\fi
\expandafter\ifx\csname citenamefont\endcsname\relax
  \def\citenamefont#1{#1}\fi
\expandafter\ifx\csname url\endcsname\relax
  \def\url#1{\texttt{#1}}\fi
\expandafter\ifx\csname urlprefix\endcsname\relax\def\urlprefix{URL }\fi
\providecommand{\bibinfo}[2]{#2}
\providecommand{\eprint}[2][]{\url{#2}}

\bibitem[{\citenamefont{Weyl}(1911)}]{Weyl1911}
\bibinfo{author}{\bibfnamefont{H.}~\bibnamefont{Weyl}},
  \bibinfo{journal}{Nachr.\ Akad.\ Wiss.\ G{\"\o}ttingen} p.
  \bibinfo{pages}{110} (\bibinfo{year}{1911}).

\bibitem[{\citenamefont{Kac}(1966)}]{Kac}
\bibinfo{author}{\bibfnamefont{M.}~\bibnamefont{Kac}}, \bibinfo{journal}{Am.\
  Math.\ Monthly} \textbf{\bibinfo{volume}{73}}, \bibinfo{pages}{1}
  (\bibinfo{year}{1966}).

\bibitem[{\citenamefont{Baltes and Hilf}(1976)}]{BaltesHilf}
\bibinfo{author}{\bibfnamefont{H.~T.} \bibnamefont{Baltes}} \bibnamefont{and}
  \bibinfo{author}{\bibfnamefont{E.~R.} \bibnamefont{Hilf}},
  \emph{\bibinfo{title}{Spectra of Finite Systems}}
  (\bibinfo{publisher}{Bibliographisches Inst. Wissenschaftsverlag, Mannheim},
  \bibinfo{year}{1976}).

\bibitem[{\citenamefont{Brack and Bhaduri}(1997)}]{BrackBhaduri}
\bibinfo{author}{\bibfnamefont{M.}~\bibnamefont{Brack}} \bibnamefont{and}
  \bibinfo{author}{\bibfnamefont{R.~K.} \bibnamefont{Bhaduri}},
  \emph{\bibinfo{title}{Semiclassical Physics}}
  (\bibinfo{publisher}{Addison-Wesley, Reading}, \bibinfo{year}{1997}).

\bibitem[{\citenamefont{Balian and Bloch}(1970)}]{BalianBloch1}
\bibinfo{author}{\bibfnamefont{R.}~\bibnamefont{Balian}} \bibnamefont{and}
  \bibinfo{author}{\bibfnamefont{C.}~\bibnamefont{Bloch}},
  \bibinfo{journal}{Ann.\ Phys.\ (N.Y.)} \textbf{\bibinfo{volume}{60}},
  \bibinfo{pages}{401} (\bibinfo{year}{1970}).

\bibitem[{\citenamefont{Balian and Bloch}(1971)}]{BalianBloch2}
\bibinfo{author}{\bibfnamefont{R.}~\bibnamefont{Balian}} \bibnamefont{and}
  \bibinfo{author}{\bibfnamefont{C.}~\bibnamefont{Bloch}},
  \bibinfo{journal}{Ann.\ Phys.\ (N.Y.)} \textbf{\bibinfo{volume}{64}},
  \bibinfo{pages}{271} (\bibinfo{year}{1971}).

\bibitem[{\citenamefont{Stewartson and Waechter}(1971)}]{SW71}
\bibinfo{author}{\bibfnamefont{K.}~\bibnamefont{Stewartson}} \bibnamefont{and}
  \bibinfo{author}{\bibfnamefont{R.~T.} \bibnamefont{Waechter}},
  \bibinfo{journal}{Proc.\ Camb.\ Phil.\ Soc.} \textbf{\bibinfo{volume}{69}},
  \bibinfo{pages}{581} (\bibinfo{year}{1971}).

\bibitem[{\citenamefont{Berry and Howls}(1994)}]{BerryHowls}
\bibinfo{author}{\bibfnamefont{M.}~\bibnamefont{Berry}} \bibnamefont{and}
  \bibinfo{author}{\bibfnamefont{C.~J.} \bibnamefont{Howls}},
  \bibinfo{journal}{Proc.\ R.\ Soc.\ Lond.\ A} \textbf{\bibinfo{volume}{447}},
  \bibinfo{pages}{527} (\bibinfo{year}{1994}).

\bibitem[{\citenamefont{Sieber et~al.}(1995)\citenamefont{Sieber, Primack,
  Smilansky, Ussishkin, and Schanz}}]{Sieber95}
\bibinfo{author}{\bibfnamefont{M.}~\bibnamefont{Sieber}},
  \bibinfo{author}{\bibfnamefont{H.}~\bibnamefont{Primack}},
  \bibinfo{author}{\bibfnamefont{U.}~\bibnamefont{Smilansky}},
  \bibinfo{author}{\bibfnamefont{I.}~\bibnamefont{Ussishkin}},
  \bibnamefont{and} \bibinfo{author}{\bibfnamefont{H.}~\bibnamefont{Schanz}},
  \bibinfo{journal}{J.\ Phys.\ A:\ Math.\ Gen.} \textbf{\bibinfo{volume}{28}},
  \bibinfo{pages}{5041} (\bibinfo{year}{1995}).

\bibitem[{\citenamefont{Couchman et~al.}(1992)\citenamefont{Couchman, Ott, and
  Antonsen}}]{Couchman92}
\bibinfo{author}{\bibfnamefont{L.}~\bibnamefont{Couchman}},
  \bibinfo{author}{\bibfnamefont{E.}~\bibnamefont{Ott}}, \bibnamefont{and}
  \bibinfo{author}{\bibfnamefont{T.~M.} \bibnamefont{Antonsen}},
  \bibinfo{journal}{Phys.\ Rev.\ A} \textbf{\bibinfo{volume}{46}},
  \bibinfo{pages}{6193} (\bibinfo{year}{1992}).

\bibitem[{\citenamefont{Prange et~al.}(1996)\citenamefont{Prange, Ott,
  Antonsen, Georgeot, and Bl{\"u}mel}}]{Prange96}
\bibinfo{author}{\bibfnamefont{R.~E.} \bibnamefont{Prange}},
  \bibinfo{author}{\bibfnamefont{E.}~\bibnamefont{Ott}},
  \bibinfo{author}{\bibfnamefont{T.~M.} \bibnamefont{Antonsen}},
  \bibinfo{author}{\bibfnamefont{B.}~\bibnamefont{Georgeot}}, \bibnamefont{and}
  \bibinfo{author}{\bibfnamefont{R.}~\bibnamefont{Bl{\"u}mel}},
  \bibinfo{journal}{Phys.\ Rev.\ E} \textbf{\bibinfo{volume}{53}},
  \bibinfo{pages}{207} (\bibinfo{year}{1996}).

\bibitem[{\citenamefont{Bl{\"u}mel
  et~al.}(1996{\natexlab{a}})\citenamefont{Bl{\"u}mel, Antonsen, Georgeot, Ott,
  and Prange}}]{Blumeletal1}
\bibinfo{author}{\bibfnamefont{R.}~\bibnamefont{Bl{\"u}mel}},
  \bibinfo{author}{\bibfnamefont{T.~M.} \bibnamefont{Antonsen}},
  \bibinfo{author}{\bibfnamefont{B.}~\bibnamefont{Georgeot}},
  \bibinfo{author}{\bibfnamefont{E.}~\bibnamefont{Ott}}, \bibnamefont{and}
  \bibinfo{author}{\bibfnamefont{R.~E.} \bibnamefont{Prange}},
  \bibinfo{journal}{Phys.\ Rev.\ Lett.} \textbf{\bibinfo{volume}{76}},
  \bibinfo{pages}{2476} (\bibinfo{year}{1996}{\natexlab{a}}).

\bibitem[{\citenamefont{Bl{\"u}mel
  et~al.}(1996{\natexlab{b}})\citenamefont{Bl{\"u}mel, Antonsen, Georgeot, Ott,
  and Prange}}]{Blumeletal2}
\bibinfo{author}{\bibfnamefont{R.}~\bibnamefont{Bl{\"u}mel}},
  \bibinfo{author}{\bibfnamefont{T.~M.} \bibnamefont{Antonsen}},
  \bibinfo{author}{\bibfnamefont{B.}~\bibnamefont{Georgeot}},
  \bibinfo{author}{\bibfnamefont{E.}~\bibnamefont{Ott}}, \bibnamefont{and}
  \bibinfo{author}{\bibfnamefont{R.~E.} \bibnamefont{Prange}},
  \bibinfo{journal}{Phys.\ Rev.\ E} \textbf{\bibinfo{volume}{53}},
  \bibinfo{pages}{3284} (\bibinfo{year}{1996}{\natexlab{b}}).

\bibitem[{\citenamefont{Kohler and
  Bl{\"u}mel}(1998{\natexlab{a}})}]{KohlerBlumel1}
\bibinfo{author}{\bibfnamefont{A.}~\bibnamefont{Kohler}} \bibnamefont{and}
  \bibinfo{author}{\bibfnamefont{R.}~\bibnamefont{Bl{\"u}mel}},
  \bibinfo{journal}{Phys.\ Lett.\ A} \textbf{\bibinfo{volume}{238}},
  \bibinfo{pages}{271} (\bibinfo{year}{1998}{\natexlab{a}}).

\bibitem[{\citenamefont{Kohler and
  Bl{\"u}mel}(1998{\natexlab{b}})}]{KohlerBlumel2}
\bibinfo{author}{\bibfnamefont{A.}~\bibnamefont{Kohler}} \bibnamefont{and}
  \bibinfo{author}{\bibfnamefont{R.}~\bibnamefont{Bl{\"u}mel}},
  \bibinfo{journal}{Phys.\ Lett.\ A} \textbf{\bibinfo{volume}{247}},
  \bibinfo{pages}{87} (\bibinfo{year}{1998}{\natexlab{b}}).

\bibitem[{\citenamefont{Kohler and
  Bl{\"u}mel}(1998{\natexlab{c}})}]{KohlerBlumel}
\bibinfo{author}{\bibfnamefont{A.}~\bibnamefont{Kohler}} \bibnamefont{and}
  \bibinfo{author}{\bibfnamefont{R.}~\bibnamefont{Bl{\"u}mel}},
  \bibinfo{journal}{Ann.\ Phys.\ (N.Y.)} \textbf{\bibinfo{volume}{267}},
  \bibinfo{pages}{249} (\bibinfo{year}{1998}{\natexlab{c}}).

\bibitem[{\citenamefont{Dabaghian et~al.}(2001)\citenamefont{Dabaghian, Jensen,
  and Bl{\"u}mel}}]{Dabaghian2001}
\bibinfo{author}{\bibfnamefont{Y.}~\bibnamefont{Dabaghian}},
  \bibinfo{author}{\bibfnamefont{R.~V.} \bibnamefont{Jensen}},
  \bibnamefont{and}
  \bibinfo{author}{\bibfnamefont{R.}~\bibnamefont{Bl{\"u}mel}},
  \bibinfo{journal}{Phys.\ Rev.\ E} \textbf{\bibinfo{volume}{63}},
  \bibinfo{pages}{066201} (\bibinfo{year}{2001}).

\bibitem[{\citenamefont{Savytskyy et~al.}(2001)\citenamefont{Savytskyy, Kohler,
  Bauch, Bl{\"u}mel, and Sirko}}]{Savytskyy2001}
\bibinfo{author}{\bibfnamefont{N.}~\bibnamefont{Savytskyy}},
  \bibinfo{author}{\bibfnamefont{A.}~\bibnamefont{Kohler}},
  \bibinfo{author}{\bibfnamefont{S.}~\bibnamefont{Bauch}},
  \bibinfo{author}{\bibfnamefont{R.}~\bibnamefont{Bl{\"u}mel}},
  \bibnamefont{and} \bibinfo{author}{\bibfnamefont{L.}~\bibnamefont{Sirko}},
  \bibinfo{journal}{Phys.\ Rev.\ E} \textbf{\bibinfo{volume}{64}},
  \bibinfo{pages}{036211} (\bibinfo{year}{2001}).

\bibitem[{\citenamefont{Bl{\"u}mel et~al.}(2002)\citenamefont{Bl{\"u}mel,
  Dabaghian, and Jensen}}]{Blumel2002}
\bibinfo{author}{\bibfnamefont{R.}~\bibnamefont{Bl{\"u}mel}},
  \bibinfo{author}{\bibfnamefont{Y.}~\bibnamefont{Dabaghian}},
  \bibnamefont{and} \bibinfo{author}{\bibfnamefont{R.~V.}
  \bibnamefont{Jensen}}, \bibinfo{journal}{Phys.\ Rev.\ E}
  \textbf{\bibinfo{volume}{65}}, \bibinfo{pages}{046222}
  (\bibinfo{year}{2002}).

\bibitem[{\citenamefont{Jones}(1986)}]{Jones86}
\bibinfo{author}{\bibfnamefont{D.~S.} \bibnamefont{Jones}},
  \emph{\bibinfo{title}{Acoustic and Electromagnetic Waves}}
  (\bibinfo{publisher}{Clarendon Press, Oxford}, \bibinfo{year}{1986}).

\bibitem[{\citenamefont{Wolfram}(1996)}]{Wolfram1996}
\bibinfo{author}{\bibfnamefont{S.}~\bibnamefont{Wolfram}},
  \emph{\bibinfo{title}{The Mathematica book}} (\bibinfo{publisher}{Cambridge
  University Press, Cambridge}, \bibinfo{year}{1996}).

\bibitem[{\citenamefont{Abramowitz and Stegun}(1965)}]{AS65}
\bibinfo{author}{\bibfnamefont{M.}~\bibnamefont{Abramowitz}} \bibnamefont{and}
  \bibinfo{author}{\bibfnamefont{I.~A.} \bibnamefont{Stegun}},
  \emph{\bibinfo{title}{Handbook of Mathematical Functions}}
  (\bibinfo{publisher}{Dover, New-York}, \bibinfo{year}{1965}).

\bibitem[{\citenamefont{Howls and Trasler}(1998)}]{HowlsTrasler1}
\bibinfo{author}{\bibfnamefont{C.~J.} \bibnamefont{Howls}} \bibnamefont{and}
  \bibinfo{author}{\bibfnamefont{S.~A.} \bibnamefont{Trasler}},
  \bibinfo{journal}{J.\ Phys.\ A:\ Math.\ Gen.} \textbf{\bibinfo{volume}{31}},
  \bibinfo{pages}{1911} (\bibinfo{year}{1998}).

\bibitem[{\citenamefont{Howls and Trasler}(1999)}]{HowlsTrasler2}
\bibinfo{author}{\bibfnamefont{C.~J.} \bibnamefont{Howls}} \bibnamefont{and}
  \bibinfo{author}{\bibfnamefont{S.~A.} \bibnamefont{Trasler}},
  \bibinfo{journal}{J.\ Phys.\ A:\ Math.\ Gen.} \textbf{\bibinfo{volume}{32}},
  \bibinfo{pages}{1487} (\bibinfo{year}{1999}).

\bibitem[{\citenamefont{Berry}(1989)}]{hyperasymp1}
\bibinfo{author}{\bibfnamefont{M.}~\bibnamefont{Berry}},
  \bibinfo{journal}{Proc.\ R.\ Soc.\ Lond.\ A} \textbf{\bibinfo{volume}{422}},
  \bibinfo{pages}{7} (\bibinfo{year}{1989}).

\bibitem[{\citenamefont{Berry and Howls}(1991)}]{hyperasymp2}
\bibinfo{author}{\bibfnamefont{M.}~\bibnamefont{Berry}} \bibnamefont{and}
  \bibinfo{author}{\bibfnamefont{C.~J.} \bibnamefont{Howls}},
  \bibinfo{journal}{Proc.\ R.\ Soc.\ Lond.\ A} \textbf{\bibinfo{volume}{434}},
  \bibinfo{pages}{657} (\bibinfo{year}{1991}).

\bibitem[{\citenamefont{Gradshteyn and Ryzhik}(1994)}]{GR5ed}
\bibinfo{author}{\bibfnamefont{I.~S.} \bibnamefont{Gradshteyn}}
  \bibnamefont{and} \bibinfo{author}{\bibfnamefont{I.~M.}
  \bibnamefont{Ryzhik}}, \emph{\bibinfo{title}{Table of Integrals, Series, and
  Products}} (\bibinfo{publisher}{Academic Press, San Diego},
  \bibinfo{year}{1994}), \bibinfo{edition}{5th} ed.

\end{thebibliography}
\end{document}